\documentclass[pra,twocolumn]{revtex4}
\usepackage{graphicx}
\begin{document}

\bibliographystyle{unsrt}

%%%%%%%%%%%%%%%%%%%%%%%%%%%%%%%%%%%%%%%%%
\title{Single Photons on Pseudo-Demand from Stored Parametric Down-Conversion}
\author{T.B. Pittman, B.C. Jacobs, and J.D. Franson}
\affiliation{The Johns Hopkins University,
Applied Physics Laboratory, Laurel, MD 20723}
%%%%%%%%%%%%%%%%%%%%%%%%%%%%%%%%%%%%%%%%%

\date{\today}

%%%%%%%%%%%%%%%%%%%%%%%%%%%%%%%%%%%%%%%%%
\begin{abstract}
We describe the results of a parametric down-conversion experiment in which the detection of one photon of a pair causes the other photon to be switched into a storage loop.  The stored photon can then be switched out of the loop at a later time chosen by the user, providing a single photon for potential use in a variety of quantum information processing applications.  Although the stored single photon is only available at periodic time intervals, those times can be chosen to match the cycle time of a quantum computer by using pulsed down-conversion. The potential use of the storage loop as a photonic quantum memory device is also discussed.
\end{abstract}
%%%%%%%%%%%%%%%%%%%%%%%%%%%%%%%%%%%%%%%%%

\maketitle

%%%%%%%%%%%%%%%%%%%%%%%%%%%%%%%%%%%%%%%%% 
\section{Introduction}
\label{sec:intro}
%%%%%%%%%%%%%%%%%%%%%%%%%%%%%%%%%%%%%%%

It is widely recognized that a reliable source of single photons on demand is required for the realization of a variety of quantum information processing applications \cite{nielsenchuangbook,bouwmeesterbook}. In the ideal case, such a source is often envisioned as a system capable of emitting a single photon at the push of a button, with spectral and temporal properties engineered in such a way \cite{branning99} that it can be used to demonstrate higher-order interference effects with photons emitted from other independent sources \cite{zukowski97}. Here we describe an alternative single-photon source for applications of this kind in which a single photon can be switched out of a storage loop at periodic time intervals.  Since these time intervals can be chosen to match the cycle time of a quantum computer, a pseudo-demand source of this kind would be just as effective as a source in which the photon can be produced at arbitrary times.

Because of the interest in single-photon sources on both a basic physics and application-driven level, searches for a practical system have been widespread.  A partial list of candidate systems includes the use of single atoms \cite{kimble77,kuhn02}, ions \cite{diedrich87}, molecules [8-11] %\cite{demartini96,kitson98,brunel99,lounis00}%
, and quantum dots [12-19] %\cite{benson00,michler00a,michler00b,zwiller01,moreau01,santori01,regelman01,yuan02}%
, as well as Coulomb-blockades \cite{kim99a}, diamond color-centers \cite{kurtsiefer00,brouri00}, microwave cavities \cite{bratke01},  quantum interference systems \cite{harris98,resch01}, and surface acoustic wave devices \cite{foden00}.  Strong evidence of photon antibunching has already been demonstrated in many of these systems, and several of them are rapidly being engineered towards the ideal of a ``push-button'' single-photon source.

In contrast to most of these sources of single photons, spontaneous parametric down-conversion is a very convenient source of pairs of photons \cite{klyshkobook}.  Because down-conversion is a coherent process subject to phase-matching conditions \cite{salehteichbook}, the photons of a chosen pair are known to be emitted in well-defined directions with respect to one another.  This aspect of down-conversion has been extremely useful for the generation of heralded single photons, in that the detection of one photon of a pair signals the presence and direction of the other photon with certainty \cite{hong86}.  Nonetheless, down-conversion has not yet been used as a source of single-photons on demand because the pairs are essentially emitted at random times \cite{franson89}. Although the use of pulsed-pump down-conversion restricts the possible pair emission time to known intervals,  the actual interval at which the pair will be emitted still cannot be chosen on demand. 

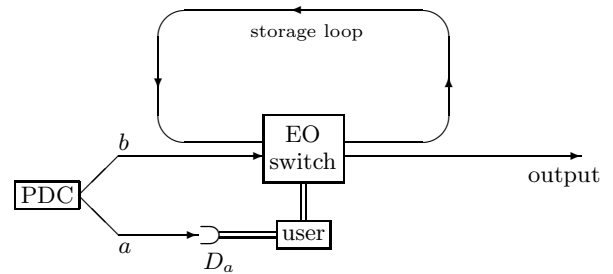
\begin{figure}[b]
\begin{center}
%%%%%%%%%%%%%%%%%%%%%%%%%%%%%%%%%%%%%%%%%%%%%%%%%%%%%%%
\begin{picture}(220,110)
%%%%% DOWN-CONVERSION SOURCE %%%%%%%%%%%
\put(6,15){\framebox(24,10){PDC}}
\put(30,20){\line(1,1){15}}
\put(45,37){$b$}
\put(30,20){\line(1,-1){15}}
\put(45,-2){$a$}
\put(45,35){\vector(1,0){55}}
\put(45,5){\vector(1,0){30}}
%%%%% DETECTOR AND CONTROL %%%%%%%%%%%
\put(75,2){\large$\supset$}
\put(77,-8){$D_{a}$}
\put(105,0){\framebox(20,10){user}}
\put(83,4){\line(1,0){22}}
\put(83,6){\line(1,0){22}}
\put(114,10){\line(0,1){15}}
\put(116,10){\line(0,1){15}}
%%%%% SWITCH %%%%%%%%%%%
\put(100,25){\framebox(30,25)}
\put(108,40){EO}
\put(102,30){switch}
\put(130,35){\vector(1,0){90}}
\put(200,25){output}
%%%%% LOOP %%%%%%%%%%%
\put(70,40){\line(1,0){30}}
\put(130,40){\line(1,0){30}}
\put(70,50){\oval(20,20)[bl]}
\put(160,50){\oval(20,20)[br]}
\put(115,80){\oval(110,20)[t]}
\put(95,80){\scriptsize storage loop}
\put(170,50){\vector(0,1){15}}
\put(170,65){\line(0,1){15}}
\put(60,80){\vector(0,-1){20}}
\put(60,60){\line(0,-1){10}}
\put(120,90){\vector(-1,0){10}}
%%%%%%%%
\end{picture}
%%%%%%%%%%%%%%%%%%%%%%%%%%%%%%%%%%%%%%%%%%%%%%%%%%%%%%%
\end{center}
\caption{A schematic overview of a source of single-photons on pseudo-demand using stored parametric down-conversion. A photon pair is emitted from a pulsed parametric down-conversion source (PDC).  Classical information describing the detection of photon $a$ by detector $D_{a}$ is fed-forward to the user (along double wires), who activates a high-speed electro-optic (EO) switch that re-routes photon $b$ into a storage loop.  Photon $b$ is then known to be circulating in the loop, and can be switched out on command by the user after any number of round trips in the ideal case.}
\label{fig:basicidea} 
\end{figure}

A potential solution to this problem is illustrated in Figure \ref{fig:basicidea}. A pair of photons is emitted by a parametric down-conversion source (PDC) pumped by a train of short pulses from, for example, a femtosecond mode-locked laser.  The pair can therefore only be emitted at one of the well-defined times defined by the repetition rate of the pulsed laser,  but the specific pulse which actually produces the pair can not be chosen in advance. Once a pair has been emitted, the detection of one of the photons is used to activate a high-speed electro-optic switch that re-routes the other photon into a storage loop.  The stored photon is then known to be circulating in the loop and can be switched out at a later time chosen by the user, providing a source of single-photons on pseudo-demand.

%%%%%%%%%%%%%%%%%% 
%%%%%%%%%%%%%%%%%% 

In this paper, we report a proof-of-principle down-conversion experiment demonstrating a single-photon source of this kind.  The current version of the experiment utilized a continuous-wave pumping laser, but an extension of the apparatus to a conventional pulsed-pump down-conversion source should be straightforward.

The remainder of the paper is outlined as follows: In section \ref{sec:experiment} we describe the details of the down-conversion experiment, which involved a 4 meter free-space storage loop along with a fiber optic delay line and high-speed Pockels cell to implement the real-time electro-optic switch. In section \ref{sec:results} we present results which clearly demonstrate the desired switch-out of single-photons after a user-chosen number of round trips through the loop.  The possibility of a practical photonic qubit memory device based on the use of this type of storage loop \cite{franson98a,franson98b}  is discussed in section \ref{sec:discussion}.  In addition, we discuss the near-term benefits of this source within the context of efficient linear optics quantum computation \cite{knill01a,franson02a}, focussing on the use of a master pulsed laser to increase the synchronization of several sources and to overcome the problems associated with false-triggering events.  A brief summary is provided in section \ref{sec:summary}.

%%%%%%%%%%%%%%%%%%%%%%%%%%%%%%%%
\section{Experiment}
\label{sec:experiment}
%%%%%%%%%%%%%%%%%%%%%%%%%%%%%%%%

%%%%%%%%%%%%%%%%%%%%%%%%%%%%%%%%
\subsection{Switching and Timing}
\label{sec:switching}
%%%%%%%%%%%%%%%%%%%%%%%%%%%%%%%%

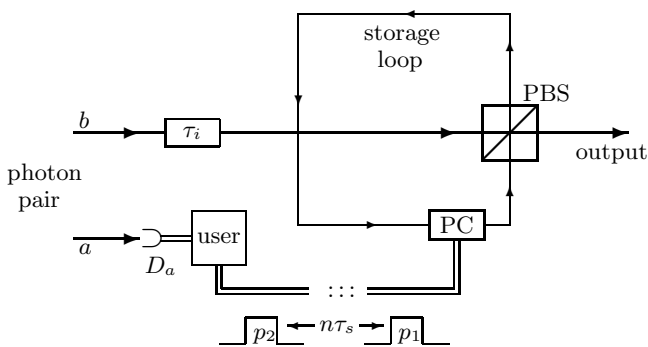
\begin{figure}[b]
\begin{center}
%%%%%%%%%%%%%%%%%%%%%%%%%%%%%%%%%%%%%%%%%%%%%%%%%%%%%%%
\begin{picture}(230,130)
%%%%% STORAGE LOOP %%%%%%%%%%%
\thicklines
\put(180,70){\framebox(20,20)}
\put(180,70){\line(1,1){20}}
\thinlines
\put(195,92){PBS}
\put(190,45){\line(0,1){80}}
\put(190,50){\vector(0,1){10}}
\put(190,105){\vector(0,1){10}}
\put(190,125){\line(-1,0){80}}
\put(160,125){\vector(-1,0){10}}
\put(110,125){\line(0,-1){80}}
\put(110,100){\vector(0,-1){10}}
\put(110,45){\line(1,0){50}}
\put(130,45){\vector(1,0){10}}
\thicklines
\put(160,40){\framebox(20,10){PC}}
\thinlines
\put(180,45){\line(1,0){10}}
\put(135,115){storage}
\put(140,105){loop}
%%%%% INPUTS %%%%%%%%%%%%%%%
\thicklines
\put(25,80){\vector(1,0){25}}
\put(27,82){$b$}
\put(50,80){\line(1,0){10}}
\put(80,80){\vector(1,0){90}}
\put(170,80){\vector(1,0){65}}
\put(25,40){\vector(1,0){25}}
\put(27,34){$a$}
\put(0,62){photon}
\put(4,52){pair}
\thinlines
\put(215,70){output}
\put(60,75){\framebox(20,10){$\tau_{i}$}}
%%%%% DETECTOR AND USER CONTROL %%%%%
\put(50,37){\large$\supset$}
\put(52,27){$D_{a}$}
\put(58,39){\line(1,0){12}}
\put(58,41){\line(1,0){12}}
\put(70,30){\framebox(20,20){user}}
\put(79,30){\line(0,-1){11}}
\put(81,30){\line(0,-1){9}}
\put(79,19){\line(1,0){35}}
\put(81,21){\line(1,0){33}}
\put(121,18){$\ldots$}
\put(121,21){$\ldots$}
\put(136,19){\line(1,0){35}}
\put(136,21){\line(1,0){33}}
\put(171,19){\line(0,1){21}}
\put(169,21){\line(0,1){19}}
%%%%% TTL PULSES %%%%%%%%%%%%%%%%
\put(80,0){\line(1,0){10}}
\put(90,0){\line(0,1){10}}
\put(90,10){\line(1,0){12}}
\put(102,10){\line(0,-1){10}}
\put(102,0){\line(1,0){10}}
\put(135,0){\line(1,0){10}}
\put(145,0){\line(0,1){10}}
\put(145,10){\line(1,0){12}}
\put(157,10){\line(0,-1){10}}
\put(157,0){\line(1,0){10}}
\put(93,2){$p_{2}$}
\put(148,2){$p_{1}$}
\put(118,5){$n\tau_{s}$}
\put(116,7){\vector(-1,0){10}}
\put(133,7){\vector(1,0){9}}
\end{picture}
%%%%%%%%%%%%%%%%%%%%%%%%%%%%%%%%%%%%%%%%%%%%%%%%%%%%%%%
\end{center}
\vspace*{-.2in}
\caption{Principle of operation of the electro-optic switch and storage loop.  The switch consists of a polarizing beamsplitter (PBS) and high-speed Pockels cell (PC) that is used to rotate the polarization of the stored photon between vertical and horizontal values. As described in the text, upon detection of the trigger photon $a$, the user sends a  classical pulse $p_{1}$ to the PC to trap photon $b$ in the storage loop.  After a chosen number of round trips, the user sends a second classical pulse $p_{2}$ to switch the stored photon out of the loop on command.}
\label{fig:switchingandtiming} 
\end{figure}

Our implementation of a high-speed electro-optic switch and storage loop is outlined in Figure \ref{fig:switchingandtiming}. The switch is comprised of a polarizing beamsplitter (PBS), which reflects vertically polarized photons and transmits horizontally polarized photons, and a Pockels cell (PC) that is used to rotate the polarization of the stored photon. The Pockels cell is configured in such a way that it will not affect the stored photon's polarization unless it is turned ``on'' by a classical pulse from the user. When it is turned on, the Pockels cell will rotate the polarization of the stored photon by $90^{o}$.

%%%%%%%%%%%%%%%%%% 
%%%%%%%%%%%%%%%%%% 

Photons $a$ and $b$ of a down-conversion pair are incident from the left, and classical information describing the detection time of photon $a$ by detector $D_{a}$ is sent forward to the user. Since the two photons of a pair are created at the same time, the user knows that photon $b$ is entering the storage loop after a fixed initial delay $\tau_{i}$.  Because photon $b$ is vertically polarized, it is reflected into the storage loop by the PBS, and would simply be reflected out of the loop and into the output mode after one round-trip if the Pockels cell were left ``off''. 

In order to store photon $b$ in the loop, the user sends a short classical pulse $p_{1}$ to the Pockels cell so that it is turned ``on'' only during the first pass of photon $b$.  The polarization of photon $b$ is therefore rotated from vertical to horizontal, causing it to be transmitted by the PBS and into the storage loop for a second round-trip. Because the Pockels cell is turned ``off'' during the second and subsequent round-trips,  photon $b$ remains horizontally polarized and continues to circulate around the storage loop.

The user can then choose when to switch out the stored photon by sending a second classical pulse, $p_{2}$.  The leading edge of $p_{2}$ is timed so that the Pockels cell rotates the polarization of photon $b$ from horizontal back to vertical after a chosen number $n$ of round trips. In this way photon $b$ is reflected out of the storage loop by the PBS and into the output mode on command.

%%%%%%%%%%%%%%%%%%%%%%%%%%%%%%%%%%
\begin{figure}[b]
\includegraphics[angle=-90,width=3.5in]{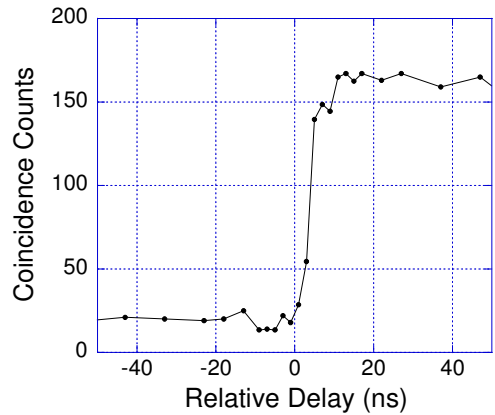}
\vspace*{-.25in}
\caption{Measurement of the Pockels cell rise time of roughly 10 ns. The data was taken by using a set-up similar to that shown in Figure \protect\ref{fig:switchingandtiming}. Additional polarizers and waveplates were used so that a down-conversion photon pair coincidence detection event could only occur if the Pockels cell was turned ``on'' at an appropriate time.  The data shows a plot of the number of coincidence counts per 60 seconds as a function of Pockels cell turn-on time delay.} 
\label{fig:risetime}
\end{figure}
%%%%%%%%%%%%%%%%%%%%%%%%%%%%%%%%%%%

Based on this principle of operation, it is clear that the round-trip time $\tau_{s}$ through the storage loop must be greater than the rise-time of the Pockels cell. As is shown in Figure \ref{fig:risetime}, the rise-time of the Pockels cell used in our experiment (ConOptics Inc. Model 360-80/D25) was measured to be roughly 10 ns, which dictated a minimum free-space storage loop length of just over 3 m.  We therefore constructed a 4 m storage loop, which allowed the stored photons to be switched out at chosen intervals every 13.3 ns.

The initial delay $\tau_{i}$ shown in Figure \ref{fig:switchingandtiming} is not part of the switching and storage loop, but was simply needed to account for the time required by the user to process the classical detection signal from $D_{a}$ and send the pulse $p_{1}$ to activate the Pockels cell.  As will be described in the next subsection, $\tau_{i}$ was on the order of 500 ns in our experiment.

%%%%%%%%%%%%%%%%%%%%%%%%%%%%%%%%
\subsection{Experimental Design}
\label{sec:experimentaldesign}
%%%%%%%%%%%%%%%%%%%%%%%%%%%%%%%%

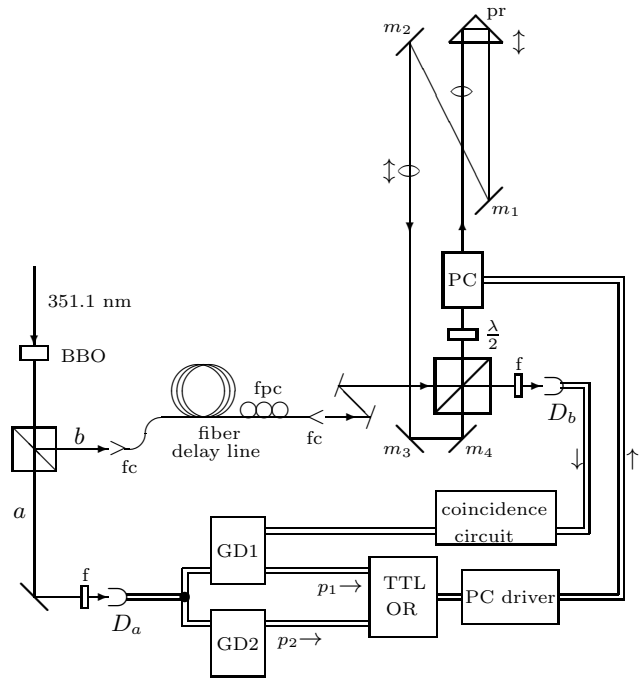
\begin{figure}[t]
\begin{center}
%%%%%%%%%%%%%%%%%%%%%%%%%%%%%%%%%%%%%%%%%%%%%%%%%%%%%%%
\begin{picture}(250,255)
%%%%% SPDC SOURCE AND INPUTS %%%%%%%%%%%%%%
\put(8,155){\vector(0,-1){30}}
\put(13,140){\scriptsize 351.1 nm}
\put(3,120){\framebox(10,5)}
\put(18,119){\scriptsize BBO}
\put(8,120){\line(0,-1){90}}
\put(8,30){\line(1,0){18}}
\put(26,26){\framebox(2,8)}
\put(26,36){\scriptsize f}
\put(28,30){\vector(1,0){8}}
\put(0,60){$a$}
\put(0,78){\framebox(16,16)}
\put(0,94){\line(1,-1){16}}
\put(8,86){\vector(1,0){28}}
\put(23,88){$b$}
%%%%% STORAGE LOOP AND OUTPUT %%%%%%%%%%%%%
\thicklines
\put(160,100){\framebox(20,20)}
\put(160,100){\line(1,1){20}}
\put(165,128){\framebox(10,3)}
\put(178,127){$\frac{\lambda}{2}$}
\put(163,140){\framebox(14,20){\scriptsize PC}}
\thinlines
\put(170,90){\line(0,1){37}}
\put(170,131){\line(0,1){9}}
\put(170,160){\line(0,1){85}}
\put(170,163){\vector(0,1){10}}
\put(150,178){\vector(0,-1){10}}
\put(165,220){$\frown$}
\put(165,218){$\smile$}
\put(170,245){\line(1,0){10}}
\put(180,245){\line(0,-1){65}}
\thicklines
\put(165,240){\line(1,0){20}}
\put(165,240){\line(1,1){10}}
\put(175,250){\line(1,-1){10}}
\put(189,238){$\updownarrow$}
\put(179,250){\scriptsize pr}
\thinlines
\put(150,240){\line(0,-1){150}}
\put(145,190){$\frown$}
\put(145,188){$\smile$}
\put(140,189){$\updownarrow$}
\put(150,240){\line(1,-2){30}}
\put(150,90){\line(1,0){20}}
\put(150,110){\line(1,0){40}}
\put(190,106){\framebox(2,8)}
\put(190,116){\scriptsize f}
\put(192,110){\vector(1,0){8}}
%%%%% MIRRORS %%%%%%%%%%%%%%%%
\thicklines
\put(3,35){\line(1,-1){10}}
\put(165,85){\line(1,1){10}}
\put(171,84){\scriptsize $m_{4}$}
\put(145,95){\line(1,-1){10}}
\put(140,84){\scriptsize $m_{3}$}
\put(145,235){\line(1,1){10}}
\put(140,244){\scriptsize $m_{2}$}
\put(175,175){\line(1,1){10}}
\put(181,175){\scriptsize $m_{1}$}
\thinlines
%%%%% FIBER %%%%%%%%%%%%%%%%%%\
\put(36,84){$>$}
\put(41,77){\scriptsize fc}
\put(42,92){\oval(16,12)[br]}
\put(58,92){\oval(16,12)[tl]}
\put(57,98){\line(1,0){55}}
\put(111,96){$<$}
\put(111,88){\scriptsize fc}
\put(70,108){\circle{18}}
\put(72,108){\circle{18}}
\put(74,108){\circle{18}}
\put(70,90){\scriptsize fiber}
\put(60,83){\scriptsize delay line}
\put(89,101){\circle{6}}
\put(95,101){\circle{6}}
\put(101,101){\circle{6}}
\put(91,107){\scriptsize fpc}
\put(118,98){\vector(1,0){12}}
\put(130,98){\line(1,0){5}}
\put(133,96){/}
\put(135,98){\line(-1,1){12}}
\put(121,108){/}
\put(123,110){\vector(1,0){35}}
%%%%% DETECTOR AND USER CONTROL %%%%%
\put(35,27){\large$\supset$}
\put(37,17){$D_{a}$}
\put(43,31){\line(1,0){22}}
\put(43,29){\line(1,0){22}}
\put(65,30){\circle*{4}}
\put(64,41){\line(0,-1){22}}
\put(66,39){\line(0,-1){18}}
\put(64,41){\line(1,0){11}}
\put(66,39){\line(1,0){9}}
\put(66,21){\line(1,0){9}}
\put(64,19){\line(1,0){11}}
\put(75,35){\framebox(20,25){\scriptsize GD1}}
\put(75,0){\framebox(20,25){\scriptsize GD2}}
\put(95,41){\line(1,0){40}}
\put(95,39){\line(1,0){40}}
\put(115,32){\scriptsize $p_{1}$}
\put(123,32){$\rightarrow$}
\put(95,21){\line(1,0){40}}
\put(95,19){\line(1,0){40}}
\put(100,12){\scriptsize $p_{2}$}
\put(108,12){$\rightarrow$}
\put(135,15){\framebox(25,30)}
\put(140,33){\scriptsize TTL}
\put(141,23){\scriptsize OR}
\put(160,31){\line(1,0){10}}
\put(160,29){\line(1,0){10}}
\put(170,20){\framebox(36,20){\scriptsize PC driver}}
\put(206,31){\line(1,0){23}}
\put(206,29){\line(1,0){25}}
\put(229,31){\line(0,1){118}}
\put(231,29){\line(0,1){122}}
\put(232,80){$\uparrow$}
\put(231,151){\line(-1,0){54}}
\put(229,149){\line(-1,0){52}}
%%%%% COINCIDENCE CIRCUIT %%%%%%%%
\put(200,107){\large$\supset$}
\put(202,97){$D_{b}$}
\put(207,111){\line(1,0){11}}
\put(207,109){\line(1,0){9}}
\put(216,109){\line(0,-1){53}}
\put(218,111){\line(0,-1){57}}
\put(211,80){$\downarrow$}
\put(216,56){\line(-1,0){11}}
\put(218,54){\line(-1,0){13}}
\put(95,56){\line(1,0){65}}
\put(95,54){\line(1,0){65}}
\put(160,50){\framebox(45,20)}
\put(162,61){\scriptsize coincidence}
\put(167,51){\scriptsize circuit}
\end{picture}
%%%%%%%%%%%%%%%%%%%%%%%%%%%%%%%%%%%%%%%%%%%%%%%%%%%%%%%
\end{center}
\vspace*{-.1in}
\caption{A schematic of the experimental set-up described in the text. In the lower portion of the figure, the detection of the trigger photon $a$ by detector $D_{a}$ is sent to two independent gate and delay generators GD1 and GD2. These units are used to produce the classical pulses $p_{1}$ and $p_{2}$ that drive a Pockels cell (PC) used to switch photon $b$ into, and out of, a 4 m free-space Gaussian transmission-line storage loop in the upper portion of the figure. Before being sent into the storage loop, photon $b$ is delayed by a fiber optic delay line for the time required to process the classical $D_{a}$ signal.}
\label{fig:experiment} 
\end{figure}

A schematic of the actual experiment is shown in Figure \ref{fig:experiment}. The set-up was a modification of our earlier experiment on feed-forward control for linear optics quantum computation, and additional technical details can be found in reference \cite{pittman02a}. The down-conversion photon pair source consisted of a 1.0 mm thick BBO crystal pumped by roughly 30 mW of the 351.1 nm line of a continuous-wave Argon-Ion laser.    
 The BBO crystal was cut for degenerate Type-II collinear phase matching and produced pairs of co-propagating, but orthogonally polarized photons at 702.2 nm \cite{rubin94}. 
The orthogonally polarized photons $a$ and $b$ of a given pair were separated using the auxiliary polarizing beamsplitter shown on the left side of Figure \ref{fig:experiment}.

%%%%%%%%%%%%%%%% 
%%%%%%%%%%%%%%%%%% 

The trigger photon $a$ passed through a 10 nm FWHM bandwidth filter, $f$,  centered at 700 nm and was detected by $D_{a}$, a Perkin Elmer model SPCM-AQR-12 single-photon counting avalanche photo-diode.
The TTL output pulse of $D_{a}$ was then sent to two independent Canberra model 410A electronic gate-and-delay generators (GD1 and GD2), which provided  the TTL pulses $p_{1}$ and $p_{2}$ described within the context of Figure \ref{fig:switchingandtiming}. The use of the two gate-and-delay generators allowed precise and simple control of the timing of the $p_{1}$ and $p_{2}$ pulses sent to the Pockels cell driver. Note that a TTL OR gate was used so that either $p_{1}$ or $p_{2}$ could be used to turn ``on'' the Pockels cell, as required. The minimum duration of $p_{1}$ and $p_{2}$ in this configuration was 100 ns, which was significantly longer than the round-trip time through the 4 m storage loop.  The timing of $p_{1}$ was therefore adjusted so that the Pockels cell transition between ``on'' and ``off'' would occur immediately after the first pass of photon $b$.  In a similar manner, the timing of $p_{2}$ was adjusted so that the transition from ``off'' back to ``on'' would occur just before the arrival of photon $b$ during its final pass through the Pockels cell.   As will be seen below, the timing of $p_{1}$ was therefore set to a fixed value relative to a detection event at $D_{a}$, while the timing of $p_{2}$ was set to several different values to demonstrate the ability to switch out the stored photon on command after a chosen number of round trips.

Before entering the storage loop, it was necessary to delay photon $b$ to account for the time required to process the detection of photon $a$ and turn on the Pockels cell with $p_{1}$.  The total required initial delay $\tau_{i}$ was therefore the sum of the operation time of several components.  The avalanche photodiode $D_{a}$ required 18 ns to produce the leading edge of its TTL output pulse \cite{pittman02a}.  Furthermore, the gate and delay generators were configured with a dead time of 200 ns, so that the production of the trailing edge of $p_{1}$ actually required 300 ns.  In addition, the TTL OR gate imposed 18 ns of delay, and the time required by the Pockels cell driver amplifiers was roughly 38 ns \cite{pittman02a}.  Various coaxial cables used to connect these devices imparted an additional 60 ns of delay. Therefore, photon $b$ had to be delayed by at least 434 ns.  As shown in Figure \ref{fig:experiment}, this was accomplished by launching photon $b$ into a 120 m  fiber optic delay line (3M brand FS-3224 single mode fiber) which provided over 500 ns of delay. Photon $b$ was launched into and out of the fiber delay line by fiber couplers (fc) comprised of suitable microscope objectives mounted on microtranslation stages.  A standard fiber polarization controller (fpc) was used to negate the effects of birefringence induced by the fiber.

After emerging from the fiber delay line, the vertically polarized photon $b$ was steered by two mirrors into the main polarizing beamsplitter and reflected into a 4 m free-space storage loop formed by a right angle prism (pr) and mirrors $m_{1}$ through $m_{4}$.  The Pockels cell was placed inside the storage loop with its fast and slow axes rotated $45^{o}$ from the vertical direction.  When we refer to the Pockels cell as being turned ``off'', it was actually DC biased in such a way that it would cause no rotation of any polarization state of photon $b$. When triggered by $p_{1}$ or $p_{2}$, the digital Pockels cell driver would turn ``on'' the Pockels cell by applying the measured half wave voltage (roughly 115 V at 702.2 nm). At these switching times, the Pockels cell would rotate the polarization of photon $b$ by $90^{o}$ as described in section \ref{sec:switching}. An additional half-wave plate ($\frac{\lambda}{2}$) was placed before the Pockels cell.  The waveplate was nominally oriented so that it did not rotate the polarization, but it could be re-oriented for various diagnostic tests.

In order to minimize beam divergence and maximize the total number of possible round trips, two 1 m focal-length lenses were used inside the storage loop.  Note that one of the lenses and the right angle prism (pr) were mounted on longitudinal translation stages.  In this way, an unfolded version of the storage loop can be viewed as a Gaussian transmission line \cite{salehteichbook} that could be optimized by their relative positions.  The degree of collimation produced by the output coupler (fc) of the fiber delay line was then adjusted in an attempt to mode-match the beam into the storage loop.  The alignment and focussing of this set-up was accomplished by coupling bright light at 694 nm from a laser diode into the fiber delay line.

After the desired number of round trips, photon $b$ was switched out of the storage loop and into the output mode, where it was detected by a second single-photon avalanche photodiode, $D_{b}$. The output of $D_{b}$, as well as an additional GD1 output signalling a $D_{a}$ event, were sent to a conventional coincidence counting circuit which, in practice, was used to record the time of arrival of the stored photon $b$ at $D_{b}$ relative to the time of arrival of the trigger photon $a$ at $D_{a}$. An accumulated histogram of the number of coincidence counts as a function of this relative time difference would therefore show a series of possible peaks separated by multiples of the 13.3 ns round-trip time of the 4 m storage loop.

%%%%%%%%%%%%%%%%%%%%%%%%%%%%%%%%
\section{Results}
\label{sec:results}
%%%%%%%%%%%%%%%%%%%%%%%%%%%%%%%%

\begin{figure}[b]
\vspace*{-.25in}
\hspace*{-.2in}
\includegraphics[angle=-90,width=3.75in]{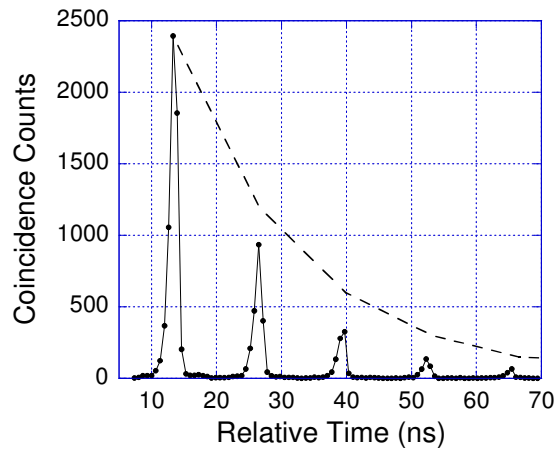}
\vspace*{-.25in}
\caption{Results of a diagnostic test to characterize the quality of the storage loop with the Pockels cell turned off and half-wave plate in the loop. The data shows a histogram of the number of coincidence counts (after 30 min.) as a function of the time of arrival of photon $b$ at $D_{b}$ relative to the time of arrival of the trigger photon $a$ at $D_{a}$.  Coincidence peaks at multiples of the 13.3 ns round-trip time are clearly seen.  The data shown here did not involve any active polarization rotation by the Pockels cell. Instead, the waveplate was used to repeatedly rotate the polarization of the stored photon to $45^{o}$, thereby giving it a 50\% chance of being switched out after each round-trip through the loop. As shown by the dashed line, an ideal lossless storage loop would therefore show a decay that goes like $(\frac{1}{2})^n$. The data was better fit by a decay of roughly $(\frac{1}{2.7})^n$ indicating a loss of roughly 26\% per round trip.  The majority of this loss was due to optical losses of the various components used to define the storage loop. }
\label{fig:decay}
\end{figure}

\begin{figure}[t]
\hspace*{-.2in}
\includegraphics[angle=-90,width=3.75in]{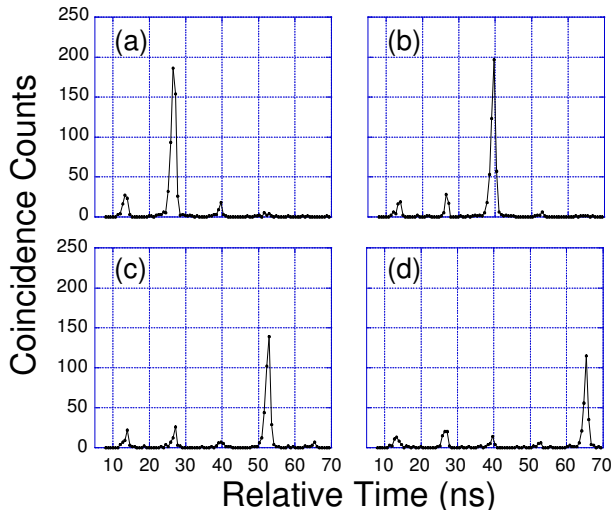}
\vspace*{-.1in}
\caption{Data demonstrating the concept of single-photons on pseudo-demand from stored parametric down-conversion. In each of the plots, the data shows coincidence count histograms (per 2 min.) analogous to that shown in Figure \protect\ref{fig:decay}. In this case, however, the half-wave plate was not used, and the real-time user control described in section \protect\ref{sec:experiment} was fully implemented.  In plot (a), the user chose to switch out the single photons after 2 round trips, while in plots (b),(c), and (d), the user chose to switch out on command after 3,4, and 5 round trips respectively.  The integrated area under the peak of interest in each of the plots is roughly consistent with the 26\% loss per round trip deduced from Figure \protect\ref{fig:decay}. In addition, small peaks resulting from switching errors described in the text can be seen in each of the plots. } 
\label{fig:peaks}
\end{figure}

Figures \ref{fig:decay} and \ref{fig:peaks} summarize the results of our proof-of-principle demonstration of a source of single-photons on pseudo-demand from stored parametric down-conversion.   Figure \ref{fig:decay} shows a histogram of the number of coincidence detections as a function of the arrival time of photon $b$ at $D_{b}$ relative to the arrival time of photon $a$ at $D_{a}$.  For this data, the Pockels cell driver was intentionally disconnected, and the half-wave plate inside the storage loop of Figure \ref{fig:experiment} was set with its fast axis rotated by $22.5^{o}$ with respect to the vertical direction. 

Upon being reflected into the storage loop by the main polarizing beamsplitter, photon $b$'s vertical polarization state was therefore rotated by $45^{o}$ (In this case the Pockels cell would do nothing but apply an overall irrelevant phase-shift for any applied voltage). After 1 round trip through the storage loop, the $45^{o}$ polarized photon $b$ therefore had a 50\% chance of being reflected out of the loop by the polarizing beamsplitter, and a 50\% chance of being transmitted back into the loop for a second round trip.  The transmitted amplitude, which was horizontally polarized, was once again rotated to $45^{o}$ by the half-wave plate, and the cycle repeated indefinitely.  

%%%%%%%%%%%%%%%%%%%%%%%%%%%%%%%%%%
%%%%%%%%%%%%%%%%%%%%%%%%%%%%%%%%%%%

The accumulated peaks shown in Figure \ref{fig:decay} therefore provide an excellent diagnostic tool for testing the quality of the storage loop and measuring the magnitude of any detrimental loop losses.  If the loop were perfectly aligned and had no optical losses, we would expect to see a coincidence count histogram that decays as $(\frac{1}{2})^{n}$, where $n$ is the number of round trips after the first pass through the loop.  This ideal case is represented by the dashed line in Figure \ref{fig:decay}. The data was better fit by a decay curve that went as $(\frac{1}{2.7})^{n}$, indicating a loss of 26\% per round trip through the loop.  Of this 26\% loss, roughly 18\% could be attributed to the various reflection and transmission losses from the optics forming the storage loop, with the remaining 8\% being due to imperfect alignment and focussing of the Gaussian transmission line.  This loop loss was lower than initially expected, and evidence of single photons completing up to twelve round trips through the loop was seen in auxiliary experiments.

%%%%%%%%%%%%%%%%%%%%%%%%%%%%%%%%%%
%%%%%%%%%%%%%%%%%%%%%%%%%%%%%%%%%%%

The main results of this paper are shown in Figures \ref{fig:peaks} (a) through (d).  The data shows coincidence count histograms analogous to that shown in Figure \ref{fig:decay}, but with the waveplate set to its nominal position to cause no polarization rotation.  In these examples, full user-control of the storage and switch-out of single-photons as described in section \ref{sec:experiment} was implemented. Figure \ref{fig:peaks} (a) shows the accumulated counts when the user decided to switch out the photon on command after 2 round trips, while Figures \ref{fig:peaks} (b), (c), and (d), respectively show switching out after choosing 3, 4, or 5 round trips. 

The integrated area under the peak of interest in each of the successive plots does decay in rough agreement with the 26\% loss per round trip deduced from Figure \ref{fig:decay}.  In addition, small undesired peaks due to switching errors can be seen in each of the four plots.  We believe that these small switching errors were primarily due to the fact that the photon beam was diverging through the Pockels cell in the storage loop, making it impossible to apply the correct half-wave voltage to the entire wavefront.  In any event, the data shown clearly demonstrates the concept of  single-photons on pseudo-demand from stored down-conversion.

%%%%%%%%%%%%%%%%%%%%%%%%%%%%%%%%
\section{Discussion}
\label{sec:discussion}
%%%%%%%%%%%%%%%%%%%%%%%%%%%%%%%%

%%%%%%%%%%%%%%%%%%%%%%%%%%%%%%%%
\subsection{False Triggering and Photon Loss Errors}
\label{sec:falsetriggering}
%%%%%%%%%%%%%%%%%%%%%%%%%%%%%%%%

In the current experiment, the effects of loop losses and switching errors were relatively small compared to the effects of false triggering events and photon loss errors. False triggering events arise from the dark-count noise events of the trigger detector $D_{a}$. These dark counts are indistinguishable from real photon $a$ detection events and cause the user to mistakenly think that a photon $b$ is stored. Similarly, photon loss errors occur when a real trigger photon $a$ is detected but the corresponding photon $b$ was lost before even entering the loop.  In practice, neither of these errors will represent a major drawback for this type of single-photon source, as described below.

In the present set-up, however, both of these errors were significant. In addition to individual detector dark counts on the order of 200 per second, the single-photon $a$ detection rate at detector $D_{a}$ was roughly 3250 counts per second, while the single-photon $b$ counting rate at detector $D_{b}$ was only 200 counts per second.  Furthermore, the observed coincidence counting rate between the two detectors was only on the order of about 10 per second. Most of this difference between the two detectors' counting rates was due to the fact that coincidence counting rates here were maximized by fully opening the apertures which defined the path to detector $D_{a}$. Since there was photon $b$ loss due to coupling inefficiency into the fiber optic delay line, as well as significant losses at a connection between two fiber patch cords comprising the delay line, this strategy helped ensure that for each successfully launched photon $b$, the twin trigger photon $a$ would not be lost. 

Although no serious efforts were made to reduce these losses and errors in our initial experiments, it should be noted that they do not pose serious problems for a practical use of this type of source.  The effects of detector dark counts can essentially be eliminated by the use of femtosecond pulsed down-conversion which allows gating of the sources and detectors.  In addition, the problems associated with photon loss errors here are of the same type as those encountered, for example, in the Klyshko absolute detector-calibration scheme \cite{klyshko80}.  Down-conversion experiments performed along those lines have shown that a near unity ratio of detection rates can be achieved (see, for example, \cite{kwiat94}). Furthermore, new techniques for efficiently coupling down-conversion radiation into optical fibers have recently been demonstrated \cite{kurtsiefer01}. 

In general, the source needs to be pumped in the conventional regime where the probability of two down-conversion pairs being emitted within the switching time is negligible.  Problems associated with low quantum efficiency values of detector $D_{a}$ are not particularly relevant; in principle, an additional switch could be used to prevent untriggered photon $b$'s from entering the storage loop. 

%%%%%%%%%%%%%%%%%%%%%%%%%%%%%%%%
\subsection{Prospects for Quantum Memory}
\label{sec:memory}
%%%%%%%%%%%%%%%%%%%%%%%%%%%%%%%%

The use of ``stopped light'' in coherently prepared atomic media has been suggested as a possible photonic memory device \cite{walsworth02}. Here we consider an alternative memory device \cite{franson98a,franson98b} using optical storage loops and switching techniques similar to those demonstrated in this paper. 

A quantum memory device must be capable of storing arbitrary qubits, or superposition states. For the case of polarization-encoded qubits, it is clear that the polarizing beamsplitter-based switch demonstrated here would be insufficient.  We have previously described a path-encoded qubit scheme for quantum memory based on the use of two phase-locked storage loops of the kind demonstrated here \cite{franson98a,franson98b}. In that scheme, the qubit values 0 and 1 are represented by two different optical paths, each of which would have its own storage loop and switch.

Despite the encouraging results obtained in our relatively simple storage loop set-up, the main technical problem in the path-encoded qubit memory scheme would be protecting the independent loops from thermally induced phase shifts. A scheme that may be more practical in the near term involves the use of polarization-encoded qubits in a single storage loop, with an interferometric-based electro-optic switch. A project involving an extension of the set-up shown in Figure \ref{fig:experiment} in this direction is currently underway in our laboratory. Quantum error correction techniques could eventually be used to greatly extend the useful memory storage time of these devices \cite{franson98a,franson98b}.

%%%%%%%%%%%%%%%%%%%%%%%%%%%%%%%%
\subsection{Use in Linear Optics Quantum Computation}
\label{sec:loqc}
%%%%%%%%%%%%%%%%%%%%%%%%%%%%%%%%

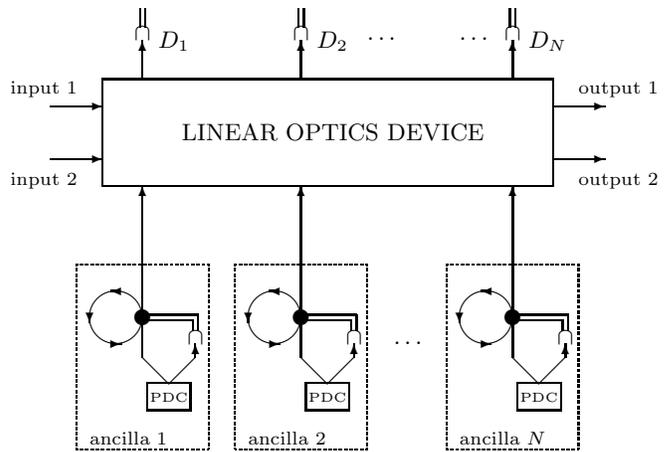
\begin{figure}[b]
\begin{center}
%%%%%%%%%%%%%%%%%%%%%%%%%%%%%%%%%%%%%%%%%
\begin{picture}(250,170)
%%%%% Inputs, Linear Optics, and Outputs %%%%%%%%%
\put(15,120){\vector(1,0){20}}
\put(0,125){\scriptsize input 1}
\put(15,100){\vector(1,0){20}}
\put(0,90){\scriptsize input 2}
\put(35,90){\framebox(170,40)}
\put(65,107){LINEAR OPTICS DEVICE}
\put(205,120){\vector(1,0){20}}
\put(215,125){\scriptsize output 1}
\put(205,100){\vector(1,0){20}}
\put(215,90){\scriptsize output 2}
%%%%% Pseudo-Demand Source 1 %%%%%%%%
\put(40,40){\circle{20}}
\put(40,50){\vector(-1,0){2}}
\put(30,40){\vector(0,-1){2}}
\put(40,30){\vector(1,0){2}}
\put(50,40){\circle*{6}}
\put(50,25){\vector(0,1){65}}
\put(60,15){\line(-1,1){10}}
\put(60,15){\line(1,1){10}}
\put(52,5){\framebox(16,10){\tiny PDC}}
\put(70,25){\vector(0,1){5}}
\put(67,30){$\cap$}
\put(69,35){\line(0,1){4}}
\put(71,35){\line(0,1){6}}
\put(69,39){\line(-1,0){19}}
\put(71,41){\line(-1,0){21}}
\put(25,-10){\dashbox(50,70)}
\put(30,-8){\scriptsize ancilla 1}
%%%%% Pseudo-Demand Source 2 %%%%%%%%
\put(100,40){\circle{20}}
\put(100,50){\vector(-1,0){2}}
\put(90,40){\vector(0,-1){2}}
\put(100,30){\vector(1,0){2}}
\put(110,40){\circle*{6}}
\put(110,25){\vector(0,1){65}}
\put(120,15){\line(-1,1){10}}
\put(120,15){\line(1,1){10}}
\put(112,5){\framebox(16,10){\tiny PDC}}
\put(130,25){\vector(0,1){5}}
\put(127,30){$\cap$}
\put(129,35){\line(0,1){4}}
\put(131,35){\line(0,1){6}}
\put(129,39){\line(-1,0){19}}
\put(131,41){\line(-1,0){21}}
\put(85,-10){\dashbox(50,70)}
\put(90,-8){\scriptsize ancilla 2}
\put(145,30){$\ldots$}
%%%%% Pseudo-Demand Source N %%%%%%%%
\put(180,40){\circle{20}}
\put(180,50){\vector(-1,0){2}}
\put(170,40){\vector(0,-1){2}}
\put(180,30){\vector(1,0){2}}
\put(190,40){\circle*{6}}
\put(190,25){\vector(0,1){65}}
\put(200,15){\line(-1,1){10}}
\put(200,15){\line(1,1){10}}
\put(192,5){\framebox(16,10){\tiny PDC}}
\put(210,25){\vector(0,1){5}}
\put(207,30){$\cap$}
\put(209,35){\line(0,1){4}}
\put(211,35){\line(0,1){6}}
\put(209,39){\line(-1,0){19}}
\put(211,41){\line(-1,0){21}}
\put(165,-10){\dashbox(50,70)}
\put(170,-8){\scriptsize ancilla $N$}
%%%%% Detector 1 %%%%%
\put(50,130){\vector(0,1){15}}
\put(47,145){$\cap$}
\put(49,150){\line(0,1){7}}
\put(51,150){\line(0,1){7}}
\put(56,142){$D_{1}$}
%%%%% Detector 2 %%%%%
\put(110,130){\vector(0,1){15}}
\put(107,145){$\cap$}
\put(109,150){\line(0,1){7}}
\put(111,150){\line(0,1){7}}
\put(116,142){$D_{2}$}
\put(135,145){$\ldots$}
\put(170,145){$\ldots$}
%%%%% Detector N %%%%%
\put(190,130){\vector(0,1){15}}
\put(187,145){$\cap$}
\put(189,150){\line(0,1){7}}
\put(191,150){\line(0,1){7}}
\put(196,142){$D_{N}$}
\end{picture}
%%%%%%%%%%%%%%%%%%%%%%%%%%%%%%%%%%%%%%%%%%%%%%%%%%%%%%%
\end{center}
\caption{A schematic illustrating the use of the single-photons on pseudo-demand from stored parametric down-conversion in efficient linear optics quantum computation \protect\cite{knill01a,franson02a}.  In addition to the two input qubits of interest on the left, $N$ ancilla photons are required as inputs to a linear optics device.  As shown by the dashed boxes, each of these $N$ inputs would be supplied by a source of the kind demonstrated in this paper.  The idea is that a master pulsed-laser would be turned on well in advance of the pre-determined gate operation time, and used for synchronized pumping of each of the $N$ sources. The pump would be turned off for each of the sources as they became occupied with a single photon in their storage loops.  All $N$ ancilla photons would then be switched into the linear optics device at the pre-determined gate operation time.}
\label{fig:loqc} 
\end{figure}

Although the single-photon source demonstrated here may be useful in a variety of quantum information applications, our primary motivation for this work was within the context of efficient linear optics quantum computation \cite{knill01a,franson02a}.  It has been shown \cite{knill01a} that near-deterministic quantum logic operations can be performed on photonic qubits using only linear optical elements, additional ancilla photons, and post-selection with feed-forward control \cite{pittman02a} based on the output of single photon detectors, as illustrated in Figure \ref{fig:loqc}. 

In addition to the qubits of interest, the input to these linear optics devices includes $N$ ancilla photons. A series of single photon detectors and feed-forward control are used to post select the desired logical output, and the operations can succeed with an error rate that scales as low as $\frac{1}{N^{2}}$ in the limit of large $N$ \cite{franson02a}.

%%%%%%%%%%%%%%%%%%%%%%%%%%%%%%%%%%%%%%%%
%%%%%%%%%%%%%%%%%%

A key requirement in this linear optics approach is clearly the need for $N$ sources of single ancilla photons.  What is required is that at some predetermined time, each of the sources will, with near certainty, emit a single photon into its assigned input mode of the linear optics device.  An approach to this problem involving the use of an array of low-power pumped down-conversion crystals at each input port is being investigated by Migdall and co-workers \cite{migdall02}. In that approach, detectors for one photon of each possible pair and fast-switching are used to select the input photon based on which crystal had successfully emitted a pair.  An approach of that kind would require a large number of efficient optical switches and down-conversion crystals for each input port of the device.

Our approach is based on the use of the single-photon source demonstrated in this paper, which essentially requires only two efficient switching events per source. The experimental results shown in Figure \ref{fig:peaks} indicate that the source of single-photons on pseudo-demand from stored down-conversion appears to be ideally suited for this task.  As shown in the lower portion of Figure \ref{fig:loqc}, one source of this kind would be used at each ancilla photon input port of the linear optics device. The entire experiment would be pumped by a master femtosecond pulsed laser which would provide the required synchronization, as well as reducing the problems associated with dark counts.  Pumping of the single-photon sources would be started well in advance of the pre-determined gate-operation time. Each of the ancilla sources would continue to be pumped until a single photon was known to be contained in its storage loop.  At the pre-determined gate-operation time, all $N$ ancilla could therefore be switched out of their loops, and into their input ports of the linear optics device.

In order to be useful in this linear optics scheme, the photons from each of the independent sources will be required to be indistinguishable in the sense that higher-order interference effects can be observed among them. At present time, the most practical method for ensuring this indistinguishability appears to be the use of  narrow-band pre-filtering techniques proposed by {\.Z}ukowski {\em et. al.} \cite{zukowski97} and used in a variety of multi-photon down-conversion experiments (see, for example \cite{bouwmeester97}).  In the example of Figure \ref{fig:loqc},  the filtering will need to be arranged in such a way that dispersion and wavepacket distortion can not be used to distinguish photons which have travelled a different number of round trips through their respective storage loops.  An alternative method for ensuring the indistinguishability of the ancilla photons would involve the source engineering techniques described by Walmsley's group \cite{walmsley01}.  The development of low-loss optical switches would also be required.  Despite these technical challenges, the single-photon source on pseudo-demand demonstrated in this paper may facilitate a near term demonstration of linear optics quantum computation.

%%%%%%%%%%%%%%%%%%%%%%%%%%%%%%%%
\section{Summary}
\label{sec:summary}
%%%%%%%%%%%%%%%%%%%%%%%%%%%%%%%%

In this paper we reported the results of an experiment that demonstrated a source of single-photons on pseudo-demand from stored parametric down-conversion.  In our experiment, the detection of one photon of a down-conversion pair was used to activate a high-speed Pockels cell that switched the other photon into a free-space storage loop.  The stored photon was then known to be circulating in the loop, and could be switched out on command after any number of round trips.  From a practical point of view, one of the attractive features of this source is that the stored photon is released into a well-defined mode as opposed to the random $4\pi$ solid-angle emission found in many single-photon sources based on spontaneous decay of a single excited two-state system.  

In the current version of our experiment, the minimum length of the storage loop was dictated by the relatively slow 10 ns rise time of the Pockels cell switch.  Although efforts could be made to reduce this time, we emphasized that the ``push-button'' aspect of any single-photon source is not the critical requirement for many quantum information applications.  What is required in many cases is simply to know, with near certainty, that a single photon will be emitted at some pre-arranged time in the future. The use of the source described in this paper with pulsed-pump parametric down-conversion appears to be an ideal candidate for these applications.

The losses and switching errors in our relatively simple proof-of-principle experiment were lower than initially expected, which is encouraging for the discussion of a photonic quantum memory device based on this kind of storage loop \cite{franson98a,franson98b}.  Much lower losses should be achievable using suitable optical fiber components. We also described how such a source of single-photons on pseudo-demand may be ideally suited as a source of ancilla photons in a linear optics quantum computation scheme \cite{knill01a,franson02a}. Useful implementations of linear optics quantum computing will require a large number of these types of sources with much smaller errors, but the results presented here provide a significant step in that direction.

%%%%%%%%%%%%%%%%%%%%%%%%%%%
\begin{center}
Acknowledgements
\end{center}
%%%%%%%%%%%%%%%%%%%%%%%%%%
This work was supported in part by ONR, ARO, NSA, ARDA, and IR\&D funding.  We would like to acknowledge useful discussions with M.J. Fitch and M. Donegan.

%%%%%%%%%%%%%%%%%%%%%%%%%%%%%%%

%%%%%%%%%%%%%%%%%%%%%%%%%%%%%%%%

\end{document}